\documentclass[10pt,conference]{IEEEtran}
\usepackage{epsfig,graphicx,subfigure,psfrag,amsmath,cases,bm}
\usepackage{latexsym,amssymb,algorithm,mathtools}
\usepackage{algorithmic}
\usepackage{color}
\usepackage{url}
\usepackage{scrtime}
\usepackage{stfloats}
\usepackage{tablefootnote}
\usepackage{cite}
\usepackage{amsfonts,mathabx}
\usepackage{textcomp}
\usepackage{xcolor,capt-of}
\usepackage{multirow}
\usepackage{amsthm}
\usepackage[left=0.625in,top=0.73in,bottom=0.95in,right=0.625in]{geometry}

\author{\textit{(Invited Paper)}\vspace*{1mm}\\
\IEEEauthorblockN {Dongfang Xu\IEEEauthorrefmark{1}, Yiming Xu\IEEEauthorrefmark{1}, Zhiqiang Wei\IEEEauthorrefmark{3}, Shenghui Song\IEEEauthorrefmark{1}, and Derrick Wing Kwan Ng\IEEEauthorrefmark{2}\\}
\IEEEauthorrefmark{1}The Hong Kong University of Science and Technology, Hong Kong;\\
\IEEEauthorrefmark{3}Xi'an Jiaotong University, China;
\IEEEauthorrefmark{2}The University of New South Wales, Australia}

\newtheorem{lemma}{Lemma}

\newtheorem{theorem}{Theorem}

\newtheorem{T-Prob}{Transformed Problem}

\DeclareMathOperator{\maxo}{maximize}
\DeclareMathOperator{\mino}{minimize}

\DeclareMathOperator{\subto}{subject\hspace*{2mm}to}

\allowdisplaybreaks

\title{Sensing-Enhanced Secure Communication: Joint Time Allocation and Beamforming Design}

\begin{document}
\maketitle
\begin{abstract}
The integration of sensing and communication enables wireless communication systems to serve environment-aware applications. In this paper, we propose to leverage sensing to enhance physical layer security (PLS) in multiuser communication systems in the presence of a suspicious target. To this end, we develop a two-phase framework to first estimate the location of the potential eavesdropper by sensing and then utilize the estimated information to enhance PLS for communication. In particular, in the first phase, a dual-functional radar and communication (DFRC) base station (BS) exploits a sensing signal to mitigate the sensing information uncertainty of the potential eavesdropper. Then, in the second phase, to facilitate joint sensing and secure communication, the DFRC BS employs beamforming and artificial noise to enhance secure communication. The design objective is to maximize the system sum rate while alleviating the information leakage by jointly optimizing the time allocation and beamforming policy. Capitalizing on monotonic optimization theory, we develop a two-layer globally optimal algorithm to reveal the performance upper bound of the considered system. Simulation results show that the proposed scheme achieves a significant sum rate gain over two baseline schemes that adopt existing techniques. Moreover, our results unveil that ISAC is a promising paradigm for enhancing secure communication in wireless networks.
\end{abstract}
\section{Introduction}
The broadcast nature of wireless information transmission is vulnerable to potential eavesdropping, thus causing a fundamental issue for wireless networks. In practice, security can be guaranteed to a certain extent by employing cryptographic encryption methods in the upper layers. However, these conventional methods may impose a heavy computational burden that inevitably introduce an unfavorable delay to wireless communication systems. To conquer these issues, advanced beamforming techniques, e.g., precoding, interference alignment, or artificial noise (AN), have been developed in recent years for physical layer security provisioning in wireless networks \cite{6781609,wu2016secure,xu2019resource,gharavol2010robust}. Yet, all these works assumed that potential eavesdroppers have been identified and their channel state information (CSI) is somehow partially known by the serving base stations (BSs), which can be overly optimistic for various practical systems. In fact, since potential eavesdroppers are usually silent and do not interact with BSs, it is challenging to detect the presence of potential eavesdroppers, let alone obtain the associated CSI. This creates a bottleneck to ensuring security in wireless networks.
\par
One promising approach to address the aforementioned issues is to integrate sensing capability in existing wireless networks. According to radar sensing theory \cite{skolnik2008radar}, by illuminating an area of interest or a target of interest with energy-focused sensing beams, we can identify the presence of an object or extract desired sensing information, e.g., angle, velocity, or distance, from the received echo signals. Inspired by this, we can deploy dual-functional radar and communication (DFRC) BSs in wireless networks to enhance physical layer security with the aid of target sensing. In the literature, some initial works have studied the beamforming design in integrated sensing and communication (ISAC) systems for security provisioning \cite{9199556,9838753,xu2022robust}. In particular, the authors of \cite{9199556} considered a multiuser ISAC system and investigated the beamforming design for the minimization of the signal-to-interference-plus-noise ratio (SINR) at a potential eavesdropper while providing satisfactory performance for communication users. Also, in \cite{xu2022robust}, the authors proposed a low-complexity robust resource allocation algorithm to maximize the sum secrecy rate of a multiuser ISAC system over a given time horizon. However, all these works focused on designing the beamforming policy with given CSI of the potential eavesdropper while ignoring the fact that the system has to first devote time and energy to obtain the CSI of the potential eavesdropper. Therefore, this calls for a comprehensive design framework that jointly optimizes the available resources. On the other hand, all the existing works, e.g., \cite{9199556,9838753,xu2022robust}, investigated only suboptimal resource allocation algorithms for secure ISAC systems. To the best of the authors' knowledge, the optimal resource allocation algorithm for secure ISAC systems is unknown in the literature, yet.
\par
Motivated by above observations, this paper proposes a two-phase design framework for ISAC systems to achieve secure communication, where the desired sensing information of the potential eavesdropper is acquired in the first phase while the secure transmission is performed in the second phase.  To this end, we jointly optimize time allocation between two phases and beamforming policy of each phase for maximizing the system sum rate while restricting the maximal information leakage. A globally optimal algorithm is developed to reveal the performance upper bound of the considered ISAC system.
\par
\textit{Notation:} Vectors and matrices are denoted by boldface lower case and boldface capital letters, respectively. The imaginary unit of a complex number is denoted by $\jmath=\sqrt{-1}$. $\mathbb{R}^{N\times M}$ and $\mathbb{C}^{N\times M}$ denote the space of $N\times M$ real-valued and complex-valued matrices, respectively. $\Re\left\{c\right\}$ and $\Im\left\{c\right\}$ represent the real part and image part of a complex number $c$, respectively. $|\cdot|$ and $||\cdot||$ denote the absolute value and the $l_2$-norm of their arguments, respectively. $(\cdot)^T$ and $(\cdot)^H$ stand for the transpose and the conjugate transpose of of their arguments, respectively. $\mathbf{I}_{N}$ refers to the identity matrix of dimension $N$. $\mathbb{H}^{N}$ denotes the set of complex Hermitian matrices of dimension $N$. $\mathrm{Tr}(\cdot)$ and $\mathrm{Rank}(\cdot)$ refer to the trace and rank of their arguments, respectively. $\mathbf{A}\succeq\mathbf{0}$ indicates that $\mathbf{A}$ is a positive semidefinite matrix. $\mathcal{CN}(0 ,\sigma^2)$ specifies the distribution of a circularly symmetric complex Gaussian (CSCG) random variable with mean $0$ and variance $\sigma^2$. $\overset{\Delta }{=}$ and $\sim$ stand for ``defined as'' and ``distributed as'', respectively. $\mathcal{E}\left \{ \cdot \right \}$ denotes statistical expectation. $\frac{\partial f}{\partial x}$ denotes the partial derivative of function $f$ with respect to variable $x$. Operation $[x]^+$ denotes $\mathrm{max}\left\{0,x\right\}$.
\par
\begin{figure}[t] 
\centering\includegraphics[width=3.0in]{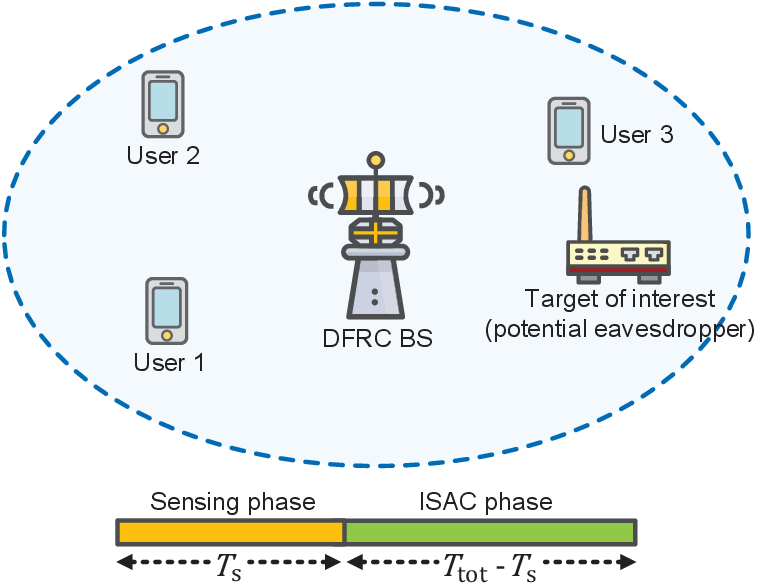}
\caption{Illustration of a multiuser ISAC system comprising one DFRC BS, $K=3$ users, and one target of interest which can be a potential eavesdropper. The scheduled time slot is divided into two orthogonal phases. The desired sensing information of the potential eavesdropper is refined in the first phase while secure information transmission is performed in the second phase.}\label{fig:1 system}
\end{figure}
\section{Multiuser ISAC System Model}
We consider a multiuser MISO ISAC system involving a DFRC BS, $K$  users, and a target of interest. The DFRC BS is equipped with a uniform linear array (ULA) consisting of $N_{\mathrm{T}}$ antennas while all the $K$ users and the target of interest are assumed to be single-antenna devices. Unlike the communication users that can collaborate with the DFRC BS, we assume that some coarse knowledge of the target (e.g., direction and distance) is available via preliminary detection of the DFRC BS \cite{9199556,xu2022robust}. Moreover, we assume that the target of interest is also a potential eavesdropper who may wiretap the information to the $K$ users.\footnote{We note that the potential eavesdropper can be identified by employing periodical scanning \cite{xu2022robust} or omnidirectional sensing beam \cite{8999605}. For consistency, we refer to the target of interest as the potential eavesdropper in the following.} To refine the sensing knowledge of the potential eavesdropper while ensuring secure communication, we focus on a scheduled time slot of duration $T_{\mathrm{tot}}$ and propose a two-phase design framework. In particular, in the first phase of duration $T_\mathrm{s}$, the DFRC BS illuminates the possible direction with a directional sensing beam based on the coarse knowledge and refines the sensing information based on the echo signal \cite{9746707}. Subsequently, in the second phase of duration $T_{\mathrm{tot}}-T_\mathrm{s}$, the DFRC BS performs beamforming to facilitate secure communication based on the refined sensing information. In this paper, we assume that the DFRC BS can obtain the perfect CSI of the communication users \cite{9199556,9838753}. As for the potential eavesdropper, the CSI uncertainty model will be discussed in Section \ref{ImperfectCSI}.
\subsection{First Phase}
During the first phase of duration $T_\mathrm{s}$, the DFRC BS transmits a probing signal $\mathbf{s}\in\mathbb{C}^{N_{\mathrm{T}}\times 1}$ to sense the potential eavesdropper. In particular, the echo signal received by the DFRC BS is given by \cite{9540344}
\begin{equation}\label{echo_signal_1}
\mathbf{y}=\Gamma\mathbf{H}\mathbf{s}+\mathbf{n},
\end{equation}
where vector $\mathbf{n}\sim\mathcal{CN}(\mathbf{0},\sigma_\mathrm{s}^2\mathbf{I}_{N_\mathrm{T}})$ is the additive white Gaussian noise (AWGN) at the DFRC BS with variance $\sigma_\mathrm{s}^2$. The variable $\Gamma\in\mathbb{C}$ indicates the coefficient of the round-trip link from the DFRC BS to the potential eavesdropper, which captures the joint effect of path loss and radar cross section (RCS) of the potential eavesdropper, and frequency of the carrier \cite{8579200}. Moreover, matrix $\mathbf{H}\in\mathbb{C}^{N_{\mathrm{T}}\times N_{\mathrm{T}}}$ denotes the target response matrix between the DFRC BS and the potential eavesdropper. In this paper, we assume that the channel between the DFRC BS and the potential eavesdropper is a line-of-sight (LoS) dominant link \cite{xu2022integrated,9199556}. Hence, $\mathbf{H}$ is defined as $\mathbf{H}=\mathbf{a}(\theta)\mathbf{a}^H(\theta)$, where $\mathbf{a}(\theta)\in\mathbb{C}^{N_\mathrm{T}\times 1}$ is the steering vector of the BS-eavesdropper link given by
\begin{equation}\label{sterringvec}
\mathbf{a}(\theta)=\Big[1,e^{\jmath2\pi\omega\mathrm{sin}\theta}, \cdots,e^{\jmath2\pi\omega(N_\mathrm{T}-1)\mathrm{sin}\theta}\Big]^T,
\end{equation}
with $\omega$ and $\theta$ representing the normalized spacing between the adjacent antenna elements and the angle of departure (AoD) from the DFRC BS to the potential eavesdropper \cite{8288677}, respectively. We note that by employing existing parameter estimation methods, e.g., \cite{32276}, \cite{1143830}, the DFRC BS is able to refine the AoD information of the potential eavesdropper based on the echo signal.
\subsection{Second Phase}
Based on the AoD of the potential eavesdropper refined in the first phase, the DFRC BS performs secure information transmission in the second phase. In particular, within the duration $T_{\mathrm{tot}}-T_\mathrm{s}$, 
the DFRC BS transmits a signal stream as follows 
\begin{equation}
    \mathbf{x}=\underset{k\in\mathcal{K}}{\sum }\mathbf{w}_kb_k+\mathbf{z}.
\end{equation}
Here, $\mathbf{w}_k\in\mathbb{C}^{N_{\mathrm{T}}\times 1}$ and $b_k\in\mathbb{C}$ indicate the beamforming vector for user $k$ and the corresponding information-bearing symbol, respectively. Without loss of generality, we assume $\mathcal{E}\big\{b_kb_k^* \big\}=1$, $\forall k \in \mathcal{K}$. Moreover, to further enhance the security while updating the sensing information of the potential eavesdropper for the next time slot, we propose to exploit AN to perform target sensing in the second phase. In particular, we model the AN signal $\mathbf{z}$ as a circularly symmetric complex Gaussian random variable, i.e.,
$\mathbf{z}\sim \mathcal{CN}\left ( \mathbf{0}, \mathbf{Z}\right )$.
Here, matrix $\mathbf{Z}\in\mathbb{H}^{N_{\mathrm{T}}}$ indicates the covariance matrix of the AN signal and it satisfies $\mathbf{Z}\succeq \mathbf{0}$.
\par
On the other hand, the received signal of user $k$ is given by
\begin{equation}
y_{\mathrm{U}_k}=\underbrace{\mathbf{g}_k^H\mathbf{w}_kb_k}_{\text{Desired signal}}+ \underbrace{\underset{\substack{j\in\mathcal{K}\setminus\left\{k\right\}}}{\sum}\mathbf{g}_k^H\mathbf{w}_jb_j}_{\text{Multiuser interference}}+\underbrace{\mathbf{g}_k^H\mathbf{z}}_{\text{Artificial noise}}+n_{\mathrm{U}_k},
\end{equation}
where vector $\mathbf{g}_k\in\mathbb{C}^{N_{\mathrm{T}}\times 1}$ represents the channel between the DFRC BS and user $k$, and scalar $n_{\mathrm{U}_k}\sim\mathcal{CN}(0,\sigma_{\mathrm{U}_k}^2)$ denotes the AWGN at user $k$ with variance $\sigma_{\mathrm{U}_k}^2$. 
\par
The received signal at the potential eavesdropper is given by
\begin{equation}
    y_{\mathrm{E}}=\mathbf{h}^H\underset{k\in\mathcal{K}}{\sum}\mathbf{w}_kb_k+\mathbf{h}^H\mathbf{z}+n_\mathrm{E},
\end{equation}
where scalar $n_\mathrm{E}\sim\mathcal{CN}(0,\sigma_{\mathrm{E}}^2)$ denotes the AWGN at the potential eavesdropper with variance $\sigma_{\mathrm{E}}^2$. Moreover, vector $\mathbf{h}\in\mathbb{H}^{N_{\mathrm{T}}}$ characterizes the LoS link from the DFRC BS to the potential eavesdropper and is given by
\begin{equation}
   \mathbf{h}=\frac{\sqrt{\beta}}{d}\mathbf{a}(\theta), 
\end{equation}
where $\beta=(\frac{\lambda_\mathrm{c}}{4\pi})^2$ 
is a constant determined by the wavelength of the adopted carrier $\lambda_\mathrm{c}$. Moreover, $d\in\mathbb{R}$ denotes the distance between the DFRC BS and the potential eavesdropper. Note that the target response matrix $\mathbf{H}$ is determined by $\mathbf{h}$ and the RCS of the target. 
\subsection{Channel Uncertainty Model}
\label{ImperfectCSI}
In this paper, we consider slowly time-varying fading channels. At the beginning of each time slot, the DFRC BS can obtain the perfect CSI of the users. However, as for the potential eavesdropper, the perfect information, e.g., AoD and distance, is challenging to obtain, due to the movement of the potential eavesdropper, hardware limitation, and estimation error. Hence, in this paper, we take into account the following uncertainties of the potential eavesdropper-involved channel when designing the ISAC system. 
\par
\textit{i) The uncertainty of AoD $\mathit{\theta}$:} First, we assume that only the coarse knowledge of the AoD $\theta$ between the DFRC BS and the potential eavesdropper is known \cite{9199556}. In particular, due to movement of the potential eavesdropper, limited beamwidth of sensing signal and/or finite angle resolution of the DFRC BS, it is challenging to obtain the accurate AoD of the potential eavesdropper. To capture the impact of $\theta$ on resource allocation design, we adopt a uncertainty model with a deterministic bound as follows
\begin{eqnarray}\label{AoDuncertainty}
    \theta=\overline{\theta}+\Delta\theta,\hspace*{2mm}
    \Omega_{\theta}\overset{\Delta }{=}\left\{\Delta\theta\hspace*{1mm}\big\vert\hspace*{1mm}\left|\Delta\theta\right|\leq \psi\right\},  
\end{eqnarray}
where $\overline{\theta}$ is the estimate of the AoD acquired when identifying the potential eavesdropper and $\Delta\theta$ denotes the AoD uncertainty. We assume that the value of the AoD uncertainty $\Delta\theta$ falls in a continuous set $\Omega_{\theta}$ while the maximum of its norm is $\psi$. Then, the steering vector in \eqref{sterringvec} is given by
\begin{equation}
\mathbf{a}(\theta)=\Big[
1,e^{\jmath2\pi\omega\mathrm{sin}(\overline{\theta}+\Delta \theta)},\hdots,e^{\jmath2\pi\omega(N_\mathrm{T}-1)\mathrm{sin}(\overline{\theta}+\Delta \theta)}\Big]^T.\label{steeringvec2}
\end{equation}
We note that $\mathbf{a}(\theta)$ is a nonlinear function with respect to $\Delta \theta$, which is challenging to handle for resource allocation algorithm design. To circumvent this difficulty, for a given estimate $\overline{\theta}$, we approximate $\mathbf{a}(\theta)$ by employing its first order Taylor series expansion as follows \cite{8974403}
\begin{equation}
\mathbf{a}(\theta)\approx \mathbf{a}^{(0)}(\overline{\theta} )+\mathbf{a}^{(1)}(\overline{\theta})(\theta-\overline{\theta} ),\label{smeq}
\end{equation}
where
\begin{eqnarray}
\hspace*{-5mm}\mathbf{a}^{(0)}(\overline{\theta} )&\hspace*{-3mm}=\hspace*{-3mm}&\begin{pmatrix}1,e^{\jmath2\pi\omega\mathrm{sin}\overline{\theta}}, \hdots,e^{\jmath2\pi\omega(N_\mathrm{T}-1)\mathrm{sin}\overline{\theta}}\end{pmatrix}^T,\\
\hspace*{-5mm}\mathbf{a}^{(1)}(\overline{\theta} )&\hspace*{-3mm}=\hspace*{-3mm}&\small\begin{pmatrix}
0,\jmath2\pi\omega\mathrm{cos}\overline{\theta},
\hdots,\jmath2\pi\left (N_\mathrm{T}-1\right )\omega\mathrm{cos}\overline{\theta}
\end{pmatrix}^T\hspace*{-1mm}\circ\mathbf{a}^{(0)}(\overline{\theta}),
\end{eqnarray}
with $\circ$ representing the Hadamard product. Then, the steering vector $\mathbf{a}(\theta)$ can be rewritten as $\mathbf{a}=\mathbf{\overline{a}}+\Delta \mathbf{a}$, where $\mathbf{\overline{a}}$ and $\Delta \mathbf{a}\in \mathbb{C}^{N_\mathrm{T}\times 1}$ are defined as
$\mathbf{\overline{a}}\overset{\Delta }{=}\mathbf{a}^{(0)}(\overline{\theta})$ and $\Delta \mathbf{a}\overset{\Delta }{=}\mathbf{a}^{(1)}(\overline{\theta} )\Delta \theta$, respectively.
\par
\textit{ii) The uncertainty of channel coefficient $\mathit{\Gamma}$:} Second, we assume that only an estimate of $\Gamma$ is known due to the clutter interference, unavoidable quantization errors, and unknown value of the RCS. As such, similar to the AoD $\theta$, we model the channel coefficient $\Gamma$ as follows
\begin{eqnarray}
\Gamma=\overline{\Gamma}+\Delta\Gamma,\hspace*{2mm}\Omega_{\Gamma}\overset{\Delta }{=}\left\{\Delta\Gamma\hspace*{1mm}\big\vert\hspace*{1mm}\left|\Delta\Gamma\right|\leq \Lambda \right\}.  
\end{eqnarray}
Here, $\overline{\Gamma}$ is the estimate of the channel coefficient acquired when identifying the potential eavesdropper and $\Delta\Gamma$ denotes the channel coefficient uncertainty. We assume that the value of the channel coefficient uncertainty $\Delta\Gamma$ falls in a continuous set $\Omega_{\Gamma}$ with the upper bound value $\Lambda$. 
\par
\textit{iii) The uncertainty of distance $\mathit{d}$:}
Third, taking into account the movement of the potential eavesdropper, unavoidable delay introduced by signal processing, and limited range resolution of the DFRC BS, we assume that only the coarse knowledge of the distance between the DFRC BS and the potential eavesdropper is known. To characterize this uncertainty, we again model the distance $d$ as follows
\begin{eqnarray}
d=\overline{d}+\Delta d,\hspace*{2mm}\Omega_{d}\overset{\Delta }{=}\left\{\Delta d\hspace*{1mm}\big\vert\hspace*{1mm}\left|\Delta d\right|\leq D \right\}.  
\end{eqnarray}
Here, $\overline{d}$ is the estimate of the distance acquired when identifying the potential eavesdropper and $\Delta d$ denotes the distance uncertainty. We assume that the value of the distance uncertainty $\Delta d$ falls in a continuous set $\Omega_{d}$ with the upper bound value $D$. 
\section{Problem Formulation}
In this section, we first define the performance metrics for the considered system. Then, we formulate the corresponding resource allocation design as an optimization problem.
\subsection{Performance Metrics}
In this paper, we adopt the Cram\'{e}r-Rao lower bound (CRLB) as the performance metric for sensing. Compared to the distance information, we are more interested in acquiring more accurate AoD information as this allows us to better exploit the spatial diversity brought by the ULA.\footnote{Due to the space limitation, we only consider the CLRB for $\theta$ in this paper. Yet, the currently adopted CRLB can be easily extended to a more general version that simultaneously takes into account $\theta$ and $d$.} 
Following the similar steps as in \cite[Appendix B]{1703855}, we derive the CRLB for estimating AoD $\theta$ as follows
\begin{equation}\label{CRLB1}
\mathrm{CRLB}(\theta)=\frac{\sigma_{\mathrm{s}}^2}{2\left| \Gamma\right|^2 T_\mathrm{s}\Big(\mathrm{Tr}\big(\frac{\partial\mathbf{H}}{\partial \theta}\mathbf{R}_\mathrm{s}\frac{\partial\mathbf{H}^H}{\partial \theta}\big)-\frac{\left|\mathrm{Tr}\big(\mathbf{H}\mathbf{R}_\mathrm{s}\frac{\partial\mathbf{H}^H}{\partial \theta}\big)\right|^2}{\mathrm{Tr}\big(\mathbf{H}\mathbf{R}_\mathrm{s}\mathbf{H}^H\big)}\Big)}.
\end{equation}
\par
To facilitate high-quality target sensing, the DFRC BS has to illuminate the directions with energy-focused beams such that the echo signal contains sufficient sensing information. Taking into account the AoD uncertainty, the direction range in the first phase is given by $\big[\overline{\theta}-\psi_{\mathrm{I}},\hspace*{1mm}\overline{\theta}+\psi_{\mathrm{I}}\big]$, where $\psi_{\mathrm{I}}$ is the maximum AoD uncertainty at the beginning of the first phase. By performing target sensing in the first phase, we can significantly reduce the AoD uncertainty to $\psi_{\mathrm{II}}$, where $\psi_{\mathrm{II}}$ is the maximum AoD uncertainty at the end of the first phase and its value is determined by the $\mathrm{CRLB}(\theta)$, cf. \eqref{CRLB1}. As such, to combat the AoD uncertainty in both phases, we discretize a given angular range into $L$ directions and generate the ideal beam pattern $\{P_p(\vartheta_l)\}_{l=1}^{L}$ for phase $p$, where $P_p(\vartheta_l)$ denotes the beam pattern power in direction $\vartheta_l$ and $p\in \left\{\mathrm{I},\mathrm{II} \right\}$. Specifically, the ideal beam pattern $\{P_p(\vartheta_l)\}_{l=1}^{L}$ of phase $p$, is given by \cite{9838753,xu2023coordinatedisac}
\begin{equation}
P_p(\vartheta_l)=\left\{\begin{matrix}
1, & \hspace*{6mm}\left|\vartheta_l -\overline{\theta}\right| \leq \left|\psi_p\right|, \\
0, & \mbox{otherwise}.\\ 
\end{matrix}\right.,\label{idealbeam1}
\end{equation}
\par
To achieve secure communication, we consider the maximum achievable
secrecy rate between the DFRC BS and user $k$, which is given by $R^{\mathrm{sec}}_k=\Big[R_k-C_k\Big]^+$. Here, $R_k$ is the achievable rate (bits/s/Hz) of user $k$ and is given by
\begin{eqnarray}
R_k=\mathrm{log}_2\Big(1+\frac{\left|\mathbf{g}_k^H\mathbf{w}_k\right|^2}{\underset{\substack{r\in\mathcal{K}\setminus\left\{k\right\}}}{\sum}\left|\mathbf{g}_k^H\mathbf{w}_r\right|^2+\mathrm{Tr}(\mathbf{g}_k\mathbf{g}_k^H\mathbf{Z})+\sigma_{\mathrm{U}_k}^2}\Big).\label{user_rate}
\end{eqnarray}
On the other hand, $C_k$ is the capacity (bits/s/Hz) of the channel between the DFRC BS and the potential eavesdropper for wiretapping the signal of user $k$ and is given by\footnote{In this paper, for security provisioning, we consider a worst-case scenario where the potential eavesdropper is equipped with unlimited computational resources and is able to cancel all the multiuser interference before decoding the information intended for a specific  user \cite{6860253}.}
\begin{equation}
C_k=\mathrm{log}_2\Big(1+\frac{\frac{\beta}{d^2}\left|\mathbf{a}^H(\theta)\mathbf{w}_k\right|^2}{\frac{\beta}{d^2}\mathrm{Tr}(\mathbf{a}(\theta)\mathbf{a}^H(\theta)\mathbf{Z})+\sigma_{\mathrm{E}}^2}\Big).
\end{equation}
\subsection{Problem Formulation}
For the system considered in this paper, we investigate the resource allocation design in a given time slot to unveil the relationship between sensing and communication. In particular, we first divide a given time slot of duration $T_{\mathrm{tot}}$ into two phases and obtain a time allocation policy $\mathcal{T}=\left\{T_{\mathrm{s}},T_{\mathrm{tot}}-T_{\mathrm{s}}\right\}$. Then, in the first phase, we aim to minimize the CRLB for estimating AoD $\theta$ by jointly designing the covariance matrix of the sensing signal and the scaling factor of the sensing beampattern $\zeta_{\mathrm{I}}$. Subsequently, in the second phase, based on the refined AoD uncertainty of the potential eavesdropper, we jointly optimize the beamforming vectors $\mathbf{w}_k$, the covariance matrix of the AN $\mathbf{Z}$, and the scaling factor of the sensing beampattern $\zeta_{\mathrm{II}}$ for the maximization the system sum rate while suppressing the information leakage to the potential eavesdropper below a pre-defined level. To this end, we formulate an optimization problem involving two subproblems as follows, 
\begin{eqnarray} 
&&\hspace*{-4mm}\underset{\substack{\mathbf{R}_\mathrm{s}\in\mathbb{H}^{N_\mathrm{T}},\\\mathbf{R}_\mathrm{s}\succeq\mathbf{0},\zeta_{\mathrm{I}}}}{\mino}\hspace*{4mm}\xi\big(\mathbf{R}_\mathrm{s}\hspace*{0.5mm}|\hspace*{0.5mm}\mathcal{T}\big)\overset{\Delta}{=}\underset{\Delta\Gamma\in\Omega_{\Gamma}}{\mathrm{max}}\hspace*{2mm}\mathrm{CRLB}(\theta)\notag\\
&&\hspace*{-4mm}\subto\hspace*{2mm}\overline{\mbox{C1}}\mbox{:}\hspace*{1mm} \mathrm{Tr}(\mathbf{R}_\mathrm{s})\leq P^{\mathrm{max}},\notag\\
&&\hspace*{15mm}\overline{\mbox{C2}}\mbox{:}\hspace*{1mm}\sum_{l=1}^{L}\left| \zeta_{\mathrm{I}} P_{\mathrm{I}}(\vartheta_l)-\mathbf{a}^H(\vartheta_l)\mathbf{R}_\mathrm{s}\mathbf{a}(\vartheta_l) \right|\leq \iota_{\mathrm{I}}.\label{prob1_1}\\
&&\hspace*{-4mm}\underset{\substack{\mathbf{Z}\in\mathbb{H}^{N_\mathrm{T}},\\\mathbf{Z}\succeq\mathbf{0},\mathbf{w}_k,\zeta_{\mathrm{II}}}}{\maxo}\hspace*{4mm}R\big(\mathbf{w}_k,\mathbf{Z}\hspace*{0.5mm}|\hspace*{0.5mm}\mathcal{T},\xi\big)\overset{\Delta}{=}\Big(1-\frac{T_\mathrm{s}}{T_{\mathrm{tot}}}\Big)\hspace*{1mm}\underset{k\in\mathcal{K} }{\sum }R_k\notag\\
&&\hspace*{-4mm}\subto\hspace*{2mm}\widetilde{\mbox{C1}}\mbox{:}\hspace*{1mm}\underset{k\in\mathcal{K} }{\sum }\left\| \mathbf{w}_k\right\|^2+\mathrm{Tr}(\mathbf{Z})\leq P^{\mathrm{max}},\notag\\
&&\hspace*{15mm}\widetilde{\mbox{C2}}\mbox{:}\hspace*{1mm}\sum_{l=1}^{L}\left| \zeta_{\mathrm{II}} P_{\mathrm{II}}(\vartheta_l)-\mathbf{a}^H(\vartheta_l)\mathbf{Z}\mathbf{a}(\vartheta_l) \right|\leq \iota_{\mathrm{II}},\notag\\
&&\hspace*{15mm}\widetilde{\mbox{C3}}\mbox{:}\hspace*{1mm}\Big(1-\frac{T_\mathrm{s}}{T_{\mathrm{tot}}}\Big)\underset{\substack{\Delta\theta\in \Omega_{\theta|\xi},\\\Delta d\in\Omega_d}}{\mathrm{max}}\hspace*{1mm}C_k\leq C_k^{\mathrm{tol}},\hspace*{1mm}\forall k.\label{prob1_2}
\end{eqnarray}
We note that the objective functions of the subproblems \eqref{prob1_1} and \eqref{prob1_2}, i.e., $\xi\big(\mathbf{R}_\mathrm{s}\hspace*{0.5mm}|\hspace*{0.5mm}\mathcal{T}\big)$ and $R\big(\mathbf{w}_k,\mathbf{Z}\hspace*{0.5mm}|\hspace*{0.5mm}\mathcal{T},\xi\big)$, are influenced by the time allocation policy $\mathcal{T}$. Also, recall that the AoD information refined in the first phase will be utilized in the second phase, the value of $\xi\big(\mathbf{R}_\mathrm{s}\hspace*{0.5mm}|\hspace*{0.5mm}\mathcal{T}\big)$ directly determines the AoD uncertainty set in \eqref{prob1_2}, i.e., $\Omega_{\theta|\xi}$, where $\Omega_{\theta|\xi}$ is the AoD uncertainty set defined in \eqref{AoDuncertainty} given the CRLB value $\xi$.\footnote{In this paper, we adopt the three-sigma rule of thumb and assume the maximum AoD uncertainty in the second phase is upper bounded by a three-fold the square root of $\mathrm{CRLB}(\theta)$, i.e., $\psi_{\mathrm{II}}=3\sqrt{\xi}$.} Considering this causality, we propose to solve \eqref{prob1_1} and \eqref{prob1_2} in a sequential manner. On the other hand, in both problems \eqref{prob1_1} and \eqref{prob1_2}, constant $P^{\mathrm{max}}$ denotes the transmit power budget of the DFRC BS. Moreover, in both phases, we employ an energy-focusing beam to illuminate the whole uncertain direction of the potential eavesdropper to update sensing information. As such, in constraints $\overline{\mbox{C2}}$ and $\widetilde{\mbox{C2}}$, we define parameters $\iota_{\mathrm{I}}$ and $\iota_{\mathrm{II}}$ to restrict the differences between the desired beam pattern and the actual beam pattern in the two phases, respectively. Besides, in problem \eqref{prob1_2}, secure communication is guaranteed by constraint $\widetilde{\mbox{C3}}$ with the lower bound of the sum $\underset{k\in\mathcal{K} }{\sum }R^{\mathrm{sec}}_k\geq\underset{k\in\mathcal{K} }{\sum }\Big[R_k-C_k^{\mathrm{tol}}\Big]^+$. Here, $C_k^{\mathrm{tol}}$ is a pre-defined parameter which is adopted to limit the maximum information leakage of user $k$.
\par
We note that optimization problem \eqref{prob1_1} is non-convex due to the quadratic term in the objective function. On the other hand, optimization problem \eqref{prob1_2} is also a non-convex problem because of the non-convex objective function and the fractional term in constraint $\widetilde{\mbox{C3}}$. Moreover, both problems are intractable as they involve continuous sets which is intractable for resource allocation design. In the literature, the optimization framework that obtain the globally optimal solution of \eqref{prob1_1} and \eqref{prob1_2} is still unknown. To fill this gap, in the next section, we develop a two-layer optimization algorithm which solves \eqref{prob1_1} and \eqref{prob1_2} optimally.
\section{Solution of the Optimization Problem}
In this section, we develop a nested-loop optimization algorithm to acquire the globally optimal solution to the formulated optimization problem. In particular, in the outer layer, we implement a line search to determine the time allocation policy. Then, for the given time allocation policy, we first solve problem \eqref{prob1_1} optimally via S-procedure. Subsequently, by capitalizing on the monotonic optimization approach, we obtain the optimal solution to \eqref{prob1_2}.
\subsection{Solution of Problem \eqref{prob1_1}}
To tackle the intractable objective function, we first recast problem \eqref{prob1_1} equivalently into its epigraph form, i.e., 
\begin{eqnarray}
\label{prob2_1}
&&\hspace*{-4mm}\underset{\substack{\mathbf{R}_\mathrm{s}\in\mathbb{H}^{N_\mathrm{T}}\\\mathbf{R}_\mathrm{s}\succeq\mathbf{0},\zeta_{\mathrm{I}},\eta}}{\mino}\hspace*{6mm}\eta\notag\\
&&\hspace*{-4mm}\subto\hspace*{4mm}\overline{\mbox{C1}}\mbox{:}\hspace*{1mm} \mathrm{Tr}(\mathbf{R}_\mathrm{s})\leq P^{\mathrm{max}},\notag\\
&&\hspace*{17mm}\overline{\mbox{C2}}\mbox{:}\hspace*{1mm}\sum_{l=1}^{L}\left| \zeta_{\mathrm{I}} P_{\mathrm{I}}(\vartheta_l)-\mathbf{a}^H(\vartheta_l)\mathbf{R}_\mathrm{s}\mathbf{a}(\vartheta_l) \right|\leq \iota_{\mathrm{I}},\notag\\
&&\hspace*{17mm}\mbox{C4:}\hspace*{1mm}\underset{\Delta\Gamma\in\Omega_{\Gamma}}{\mathrm{max}}\hspace*{2mm}\mathrm{CRLB}(\theta)\leq \eta,
\end{eqnarray}
where $\eta\in\mathbb{R}$ is a slack variable. Then, substituting \eqref{CRLB1} into constraint C4 yields
\begin{equation}
    \mbox{C4:}\hspace*{1mm}\frac{1}{\Big(\mathrm{Tr}\big(\frac{\partial\mathbf{H}}{\partial \theta}\mathbf{R}_\mathrm{s}\frac{\partial\mathbf{H}^H}{\partial \theta}\big)-\frac{\left|\mathrm{Tr}\big(\mathbf{H}\mathbf{R}_\mathrm{s}\frac{\partial\mathbf{H}^H}{\partial \theta}\big)\right|^2}{\mathrm{Tr}\big(\mathbf{H}\mathbf{R}_\mathrm{s}\mathbf{H}^H\big)}\Big)}\leq \underset{\Delta\Gamma\in\Omega_{\Gamma}}{\mathrm{min}}\hspace*{1mm}\frac{2}{\sigma_{\mathrm{s}}^2}\left| \Gamma\right|^2\eta.
\end{equation}
For ease of handling the uncertainty in constraint C4, we further define an auxiliary optimization variable $\chi\in\mathbb{R}$ and decompose constraint C4 equivalently into the following two constraints
\begin{eqnarray}
    &&\hspace*{-4mm}\mbox{C4a:}\hspace*{1mm}\mathrm{Tr}\big(\frac{\partial\mathbf{H}}{\partial \theta}\mathbf{R}_\mathrm{s}\frac{\partial\mathbf{H}^H}{\partial \theta}\big)-\frac{\left|\mathrm{Tr}\big(\mathbf{H}\mathbf{R}_\mathrm{s}\frac{\partial\mathbf{H}^H}{\partial \theta}\big)\right|^2}{\mathrm{Tr}\big(\mathbf{H}\mathbf{R}_\mathrm{s}\mathbf{H}^H\big)}\geq \chi,\\
    &&\hspace*{-4mm}\mbox{C4b:}\hspace*{1mm}\frac{1}{\chi}\leq \underset{\Delta\Gamma\in\Omega_{\Gamma}}{\mathrm{min}}\hspace*{2mm}\frac{2}{\sigma_{\mathrm{s}}^2}\left| \Gamma\right|^2\eta.
\end{eqnarray}
Then, by exploiting the Schur complement, we can further rewrite constraint C4a equivalently as the following convex constraint \cite{boyd2004convex,xunetworkisac}
\begin{equation}
    \overline{\mbox{C4a}}\mbox{:}\hspace*{1mm}\begin{bmatrix}
\mathrm{Tr}\big(\frac{\partial\mathbf{H}}{\partial \theta}\mathbf{R}_\mathrm{s}\frac{\partial\mathbf{H}^H}{\partial \theta}\big)-\chi & \mathrm{Tr}\big(\mathbf{H}\mathbf{R}_\mathrm{s}\frac{\partial\mathbf{H}^H}{\partial \theta}\big) \\
\hspace*{-9mm}\mathrm{Tr}\big(\mathbf{H}\mathbf{R}_\mathrm{s}\frac{\partial\mathbf{H}^H}{\partial \theta}\big) &  \mathrm{Tr}\big(\mathbf{H}\mathbf{R}_\mathrm{s}\mathbf{H}^H\big)\\
\end{bmatrix}\succeq \mathbf{0}.
\end{equation}
\par
Next, we tackle the semi-infinite constraint C4b by applying the following lemma.
\begin{lemma}
\textit{(S-Procedure \cite{boyd2004convex}}) Two functions $u_i(\mathbf{c}): \mathbb{C}^{N\times 1}\to \mathbb{R}$, $i\in \left \{ 1,2 \right \}$, with vector variable $\mathbf{c}\in \mathbb{C}^{N\times 1}$, are defined as
\begin{equation}
u_i(\mathbf{c})= \mathbf{c}^H\mathbf{Q}_i\mathbf{c}+2\Re\left \{\mathbf{q}^H_i\mathbf{c}  \right \}+q_i,
\end{equation}
where $\mathbf{Q}_i\in \mathbb{H}^N$, $\mathbf{q}_i\in \mathbb{C}^{N\times 1}$, and $\mathrm{q}_i\in \mathbb{R}$. Then, we can ensure the implication $u_1(\mathbf{c})\leq0 \Rightarrow u_2(\mathbf{c})\leq0$ if there exists a non-negative scalar $\delta \geq 0$ satisfying the following linear matrix inequality (LMI)
\begin{equation}
\delta\begin{bmatrix}
\mathbf{Q}_1 &  \mathbf{q}_1\\
\mathbf{q}_1^H &  \mathit{q}_1
\end{bmatrix}-\begin{bmatrix}
\mathbf{Q}_2 &  \mathbf{q}_2\\
\mathbf{q}_2^H &  \mathit{q}_2
\end{bmatrix}\succeq \mathbf{0}.
\end{equation}
\end{lemma}
\par
To facilitate the application of S-procedure, we rewrite constraint C4b as follows
\begin{equation}
    \mbox{C4b}\mbox{:}\hspace*{1mm}\frac{1}{\chi}-\frac{2\eta}{\sigma_{\mathrm{s}}^2}\big((\Delta\Gamma)^2+2\Delta\Gamma\overline{\Gamma}+\overline{\Gamma}^2\big)\leq 0.
\end{equation}
Then, utilizing \textbf{Lemma 1}, the following implication can be obtained: $(\Delta\Gamma)^2-\Lambda^2\leq 0 \Rightarrow$ C4b holds if and only if there exists $\delta\geq 0$ satisfying the following LMI constraint $\overline{\mbox{C4b}}$
\begin{equation}
   \overline{\mbox{C4b}}\mbox{:}\hspace*{1mm}\delta \begin{bmatrix}
1 &  0\\
0 &  -\Lambda^2\\
\end{bmatrix}+  \begin{bmatrix}
\frac{2\eta}{\sigma_{\mathrm{s}}^2} &  \frac{4\eta}{\sigma_{\mathrm{s}}^2}\overline{\Gamma}\\
\frac{4\eta}{\sigma_{\mathrm{s}}^2}\overline{\Gamma} &  \frac{4\eta}{\sigma_{\mathrm{s}}^2}\overline{\Gamma}^2-\frac{1}{\chi}\\
\end{bmatrix}\succeq\mathbf{0}.
\end{equation}
\par
Now, problem \eqref{prob1_1} is reformulated equivalently as the following convex optimization problem \cite{ben2009robust}
\begin{eqnarray}
\label{prob3_1}
&&\hspace*{-4mm}\underset{\substack{\mathbf{R}_\mathrm{s}\in\mathbb{H}^{N_\mathrm{T}},\mathbf{R}_\mathrm{s}\succeq\mathbf{0},\\\zeta_{\mathrm{I}},\eta,\chi,\delta}}{\mino}\hspace*{6mm}\eta\notag\\
    &&\hspace*{-2mm}\subto\hspace*{6mm}\overline{\mbox{C1}},\overline{\mbox{C2}},\overline{\mbox{C4a}},\overline{\mbox{C4b}}.
\end{eqnarray}
We note that problem \eqref{prob3_1} can be efficiently and optimally solved by a standard convex program solver such as CVX \cite{grant2008cvx}.
\subsection{Solution of Problem \eqref{prob1_2}}
To facilitate  the optimal algorithm design, we define beamforming matrix $\mathbf{W}_k\in\mathbb{C}^{N_{\mathrm{T}}\times N_{\mathrm{T}}}$ as $\mathbf{W}_k=\mathbf{w}_k\mathbf{w}_k^H$, $\forall k$. Then, we rewrite problem \eqref{prob2_1} equivalently as follows
\begin{eqnarray}
\label{prob2_2}
    &&\hspace*{-4mm}\underset{\substack{\mathbf{W}_k,\mathbf{Z}\in\mathbb{H}^{N_\mathrm{T}},\\\mathbf{W}_k,\mathbf{Z}\succeq\mathbf{0},\zeta_{\mathrm{II}}}}{\maxo}\hspace*{3mm}\Big(1-\frac{T}{T_{\mathrm{tot}}}\Big)\hspace*{2mm}\underset{k\in\mathcal{K} }{\sum }\mathrm{log}_2\Big(1+\mathrm{SINR}_k\Big)\notag\\
    &&\hspace*{-4mm}\subto\hspace*{2mm}\widetilde{\mbox{C1}}\mbox{:}\hspace*{1mm}\underset{k\in\mathcal{K} }{\sum }\mathrm{Tr}(\mathbf{W}_k)+\mathrm{Tr}(\mathbf{Z})\leq P^{\mathrm{max}},\notag\\
    &&\hspace*{15mm}\widetilde{\mbox{C2}}\mbox{:}\hspace*{1mm}\sum_{l=1}^{L}\left| \zeta_{\mathrm{II}} P_{\mathrm{II}}(\vartheta_l)-\mathbf{a}^H(\vartheta_l)\mathbf{Z}\mathbf{a}(\vartheta_l) \right|\leq \iota_{\mathrm{II}},\notag\\
    &&\hspace*{15mm}\widetilde{\mbox{C3}}\mbox{:}\hspace*{1mm}\underset{\substack{\Delta\theta\in \Omega_{\theta|\xi},\\\Delta d\in\Omega_d}}{\mathrm{max}}\hspace*{2mm}C_k\leq\overline{C}_k^{\mathrm{tol}},\hspace*{1mm}\forall k,\notag\\
    &&\hspace*{15mm}\widetilde{\mbox{C5}}\mbox{:}\hspace*{1mm}\mathrm{Rank}(\mathbf{W}_k)\leq 1, \hspace*{1mm}\forall k,
\end{eqnarray}
where scalars $\mathrm{SINR}_k$ and $\overline{C}_k^{\mathrm{tol}}$ are defined by $ \mathrm{SINR}_k=\frac{\mathrm{Tr}\big(\mathbf{G}_k\mathbf{W}_k\big)}{\underset{\substack{r\in\mathcal{K}\setminus\left\{k\right\}}}{\sum}\mathrm{Tr}\big(\mathbf{G}_k\mathbf{W}_r\big)+\mathrm{Tr}(\mathbf{G}_k\mathbf{Z})+\sigma_{\mathrm{U}_k}^2}$ and $\overline{C}_k^{\mathrm{tol}}=C_k^{\mathrm{tol}}(\frac{T_{\mathrm{tot}}}{T_{\mathrm{tot}}-T_\mathrm{s}})$, respectively. Also, $\mathbf{W}_k\in\mathbb{H}^{N_\mathrm{T}}$, $\mathbf{W}_k\succeq\mathbf{0}$, and the rank-one constraint $\widetilde{\mbox{C5}}$ compel the solution of \eqref{prob2_2} to be feasible to the original problem \eqref{prob1_2}. Next, to convexify the semi-infinite constraint $\widetilde{\mbox{C3}}$, we first explicitly express it as follows
\begin{equation}\label{C3_orig}
\hspace*{-1mm}\widetilde{\mbox{C3}}\mbox{:}\hspace*{1mm}\underset{\substack{\Delta\theta\in \Omega_{\theta|\xi},\\\Delta d\in\Omega_d}}{\mathrm{max}}\hspace*{2mm}\frac{\frac{\beta}{d^2}\mathrm{Tr}\big(\mathbf{a}(\theta)\mathbf{a}^H(\theta)\mathbf{W}_k\big)}{\frac{\beta}{d^2}\mathrm{Tr}\big(\mathbf{a}(\theta)\mathbf{a}^H(\theta)\mathbf{Z}\big)+\sigma_{\mathrm{E}}^2}\leq2^{\overline{C}_k^{\mathrm{tol}}}\hspace*{-0.5mm}-\hspace*{-0.5mm}1,\hspace*{1mm}\forall k.
\end{equation}
Then, we divide the denominator and numerator of the fractional term of \eqref{C3_orig} with $\frac{\beta}{d^2}$ and introduce a slack variable $\tau_k\in\mathbb{R}$ to decompose constraint $\widetilde{\mbox{C3}}$ into the following a pair of constraints \cite{8974403}
\begin{eqnarray}
&&\hspace*{-11mm}\widetilde{\mbox{C3a}}\mbox{:}\hspace*{0mm}\underset{\Delta\theta\in \Omega_{\theta|\xi}}{\mathrm{max}}\hspace*{-1mm}\mathrm{Tr}\big(\mathbf{a}(\theta)\mathbf{a}^H\hspace*{-0.5mm}(\theta)\mathbf{W}_k\big)-\overline{c}\mathrm{Tr}\big(\mathbf{a}(\theta)\mathbf{a}^H\hspace*{-0.5mm}(\theta)\mathbf{Z}\big)\hspace*{-0.5mm}\leq\hspace*{-0.5mm}\tau_k,\\
&&\hspace*{-11mm}\widetilde{\mbox{C3b}}\mbox{:}\hspace*{1mm}\tau_k\beta\leq\underset{\Delta d\in\Omega_d}{\mathrm{min}}\hspace*{2mm}\overline{c}d^2\sigma_{\mathrm{E}}^2,
\end{eqnarray}
where constant $\overline{c}$ is given by $\overline{c}=2^{\overline{C}_k^{\mathrm{tol}}}-1$. Subsequently, similar to constraint C4b, we employ S-procedure to transform constraints C3a and C3b into the following LMI constraints
\begin{eqnarray}
&&\hspace*{-4mm}\widetilde{\widetilde{\mbox{C3a}}}\mbox{:}\hspace{1mm}\begin{bmatrix}
\kappa_k & 0\notag\\
0 & \tau_k-\kappa_k\psi^2\end{bmatrix}-\mathbf{U}^H_{\mathbf{a}}[\mathbf{W}_k-\overline{c}\mathbf{Z}]\mathbf{U}_{\mathbf{a}}\succeq \mathbf{0},\hspace*{1mm}\forall k,\label{SC2a}\\
&&\hspace*{-4mm}\widetilde{\widetilde{\mbox{C3b}}}\mbox{:}\hspace*{1mm}\begin{bmatrix}
\varrho_k+\widetilde{\sigma} & \overline{c}\overline{d}\sigma_{\mathrm{E}}^2\\
\overline{c}\overline{d}\sigma_{\mathrm{E}}^2 & -\varrho_kD^2-\tau_k\beta+\overline{c}\overline{d}^2\sigma_{\mathrm{E}}^2
\end{bmatrix}\succeq\mathbf{0},\hspace*{1mm}\forall k,\label{SC2b}
\end{eqnarray}
where matrix $\mathbf{U}_{\mathbf{a}}\in\mathbb{C}^{N_\mathrm{T} \times 2}$ is given by $\mathbf{U}_{\mathbf{a}}=\left [ \mathbf{a}^{(1)}(\overline{\theta})\hspace*{2mm}\mathbf{\overline{a}} \right ]$ and variables $\kappa_k$, $\varrho_k\geq 0$.
\begin{algorithm}[t]
\caption{Polyblock Approximation Algorithm}
\begin{algorithmic}[1]
\small
\STATE Construct polyblock $\mathcal{P}^{(1)} $ with vertex set $\mathcal{V}^{(1)}=\left\{\mathbf{v}^{(1)} \right\}$, set convergence tolerance factor $0\leq\epsilon_{\mathrm{PA}}\ll 1$ and iteration index $m=1$
\REPEAT
\STATE  Calculate the projection of vertex $\mathbf{v}^{(m)}$ onto set $\mathcal{F}$, i.e., $\bm{\pi}(\mathbf{v}^{(m)})$, via \textbf{Algorithm 2}
\STATE Generate a set $\widehat{\mathcal{V}}^{(m)}$ that contains $K$ new vertices, i.e., $\widehat{\mathcal{V}}^{(m)}=\left\{\widehat{\mathbf{v}}_1^{(m)},\cdots,\widehat{\mathbf{v}}_K^{(m)}\right\}$, where $\widehat{\mathbf{v}}_i^{(m)}=\mathbf{v}^{(m)}-\big(v^{(m)}_k-\pi_k(\mathbf{v}^{(m)})\big)\mathbf{e}_k$, $\forall k\in\mathcal{K}$
\STATE Construct a smaller polyblock $\mathcal{P}^{(m+1)}$ with new vertex set $\mathcal{V}^{(m+1)}= \big(\mathcal{V}^{(m)}\setminus\left\{\mathbf{v}^{(m)}\right\} \big)\cup \widehat{\mathcal{V}}^{(m)}$
\STATE Find $\mathbf{v}^{(m+1)}$ from $\mathcal{V}^{(m+1)}$ whose projection maximizes the objective function of the problem, i.e., $\mathbf{v}^{(m+1)}=\underset{\mathbf{v}\in\mathcal{V}^{(m+1)}}{\mathrm{arg\hspace*{1mm}max\hspace*{1mm}}}\Phi\big(\bm{\pi}(\mathbf{v}^{(m+1)})\big)$.
\STATE Set $m=m+1$
\UNTIL $\left |\Phi\big(\mathbf{v}^{(m)}\big)-\Phi\big(\bm{\pi}(\mathbf{v}^{(m)})\big)\right |\leq\epsilon _{\mathrm{PA}}$
\STATE Obtain the optimal solution $\mathbf{W}^*_k$, $\mathbf{Z}^*$, and $\zeta_{\mathrm{II}}^*$
\label{Algorithm 1}
\end{algorithmic}
\end{algorithm}
\par
Next, we exploit monotonic optimization theory \cite{tuy2000monotonic} to find the optimal solution of \eqref{prob2_2}. To start with, we introduce a slack variable $\bm{\mu}\in\mathcal{F}$, where set $\mathcal{F}\in\mathbb{R}^K$ is given by
\begin{equation}
\hspace*{-2mm}\mathcal{F}=\left \{\hspace*{1mm}\bm{\mu}\hspace*{1mm}\Big\vert\hspace*{1mm}1\leq \mu_k\leq \frac{f_k(\mathbf{W},\mathbf{Z})}{g_k(\mathbf{W},\mathbf{Z})}, \hspace*{0.5mm}(\mathbf{W},\mathbf{Z})\in\mathcal{G},\hspace*{0.5mm}\forall k \in \mathcal{K}\right\}.
\end{equation}
Here, variable $\mathbf{W}$ is the collection of all the matrices $\mathbf{W}_k$. Moreover, functions $f_k(\mathbf{W},\mathbf{Z})$ and $g_k(\mathbf{W},\mathbf{Z})$ are given by, respectively,
\begin{eqnarray}
    &&\hspace*{-8mm}f_k(\mathbf{W},\mathbf{Z})=\underset{j\in\mathcal{K}}{\sum}\mathrm{Tr}\big(\mathbf{G}_k\mathbf{W}_j\big)+\mathrm{Tr}(\mathbf{G}_k\mathbf{Z})+\sigma_{\mathrm{U}_k}^2,\\
    &&\hspace*{-8mm}g_k(\mathbf{W},\mathbf{Z})=\underset{\substack{r\in\mathcal{K} \setminus\left\{k\right\}}}{\sum}\mathrm{Tr}\big(\mathbf{G}_k\mathbf{W}_r\big)+\mathrm{Tr}(\mathbf{G}_k\mathbf{Z})+\sigma_{\mathrm{U}_k}^2,
\end{eqnarray}
and set $\mathcal{G}$ is established by constraints $\widetilde{\mbox{C1}}$, $\widetilde{\mbox{C2}}$, $\widetilde{\widetilde{\mbox{C3a}}}$, $\widetilde{\widetilde{\mbox{C3b}}}$, and $\widetilde{\mbox{C5}}$.
\par
Now, the optimization problem \eqref{prob2_2} is rewritten as the following canonical monotonic optimization problem
\begin{eqnarray}
\label{prob2_3}
    &&\hspace*{-4mm}\underset{\substack{\mathbf{W}_k,\mathbf{Z}\in\mathbb{H}^{N_\mathrm{T}},\\\mathbf{W}_k,\mathbf{Z}\succeq\mathbf{0},\zeta_{\mathrm{II}},\\\tau_k,\kappa_k,\varrho_k,\mu_k}}{\maxo}\hspace*{4mm}\Phi(\bm{\mu}) = \underset{k\in\mathcal{K} }{\sum }\mathrm{log_2}(\mu_k)\notag\\
    &&\hspace*{-4mm}\subto\hspace*{4mm}\bm{\mu}\in\mathcal{F}.
\end{eqnarray}
We note that the objective function of problem \eqref{prob2_3} is a monotonic increasing function in $\mathbf{\mu}$ over the feasible set $\mathcal{F}$. Hence, the optimal solution, referred to as $\mathbf{\mu}^*$, lies on the upper boundary of the feasible set $\mathcal{F}$. Next, be employing the polyblock approximation \cite{tuy2000monotonic}, we design a globally optimal algorithm for the monotonic optimization problem \eqref{prob2_3}. Considering the fact that the upper boundary of $\mathcal{F}$ is usually unknown, we propose to approach the boundary by iteratively pruning a pre-defined polyblock $\mathcal{P}^{(1)}$ while ensuring the resulting polyblock always contains the feasible set. To start with, we generate one vertex $\mathbf{v}^{(1)}\in\mathbb{R}^K$ associated with the vertex set $\mathcal{V}^{(1)}$ and construct the corresponding polyblock $\mathcal{P}^{(1)}$ that enfolds the feasible set $\mathcal{F}$.\footnote{The generation of the first vertex and the initialization will be discussed later.} Based on the vertex $\mathbf{v}^{(1)}$, we generate $K$ new vertices in the vertex set $\widehat{\mathcal{V}}^{(1)}=\left\{\widehat{\mathbf{v}}_1^{(1)},\cdots,\widehat{\mathbf{v}}_{K}^{(1)}\right\}$. Specifically, we calculate $\widehat{\mathbf{v}}_k^{(1)}=\mathbf{v}^{(1)}-(v^{(1)}_k-\pi_k(\mathbf{v}^{(1)}))\mathbf{e}_k$, $\forall k\in\mathcal{K}$, where scalar variables $v^{(1)}_k$ and $\pi_k(\mathbf{v}^{(1)})$ denote the $k$-th elements of $\mathbf{v}^{(1)}$ and $\bm{\pi}(\mathbf{v}^{(1)})$, respectively. Here, $\bm{\pi}(\mathbf{v}^{(1)})\in \mathbb{R}^K$ represents the projection of the vertex $\mathbf{v}^{(1)}$ onto the upper boundary of set $\mathcal{F}$ and $\mathbf{e}_k$ is a unit vector with element $k$ equal to 1. Subsequently, the polyblock $\mathcal{P}^{(1)}$ is shrunken by substituting the vertex $\mathbf{v}^{(1)}$ with $K$ new vertices in $\widehat{\mathcal{V}}^{(1)}$ and the new polyblock $\mathcal{P}^{(2)}$ still satisfies $\mathcal{P}^{(2)}\supset\mathcal{F}$. Accordingly, we update the vertex set of $\mathcal{P}^{(2)}$ as $\mathcal{V}^{(2)}=\widehat{\mathcal{V}}^{(1)} \bigcup\big(\mathcal{V}^{(1)}\setminus\left \{\mathbf{v}^{(1)} \right \}\big)$. Then, for each vertex in the set $\mathcal{V}^{(2)}$, we compute the projection onto the upper boundary of $\mathcal{F}$. Afterward, the vertex whose projection maximizes the objective function of problem \eqref{prob2_3} is selected as the optimal vertex $\mathbf{v}^{(2)}$ in $\mathcal{V}^{(2)}$, i.e., $\mathbf{v}^{(2)}=\underset{\mathbf{v}\in\mathcal{V}^{(2)}}{\mathrm{arg\hspace*{1mm}max\hspace*{1mm}}}\Phi\big(\bm{\pi}(\mathbf{v}^{(2)})\big)$. The above steps are implemented again to shrink $\mathcal{P}^{(2)}$ based on vertex $\mathbf{v}^{(2)}$. As a result, we keep obtaining a smaller polyblock during iterations, resulting in a sequence of polyblocks satisfying $\mathcal{P}^{(1)}\supset\mathcal{P}^{(2)} \supset\cdots\supset\mathcal{F}$. The algorithm terminates if $\left |\Phi\big(\mathbf{v}^{(m)}\big)-\Phi\big(\bm{\pi}(\mathbf{v}^{(m)})\big)\right |\leq\epsilon _{\mathrm{PA}}$, where $m$ is the iteration index and constant $\epsilon_{\mathrm{PA}}>0$ is the pre-defined convergence tolerance factor which is adopted to specify the accuracy of the algorithm. The key steps of the developed polyblock approximation algorithm are summarized in \textbf{Algorithm 1}.
\par
\begin{algorithm}[t]
\caption{Projection Bisection Search Algorithm}
\begin{algorithmic}[1]
\small
\STATE Initialize $\varpi_{\mathrm{min}}=0$, $\varpi_{\mathrm{max}}=1$, iteration index $j=1$, and convergence tolerance $0<\delta_{\mathrm{BS}} \ll 1$
\REPEAT
\STATE Calculate $\varpi_j$ based on $\varpi_j=(\varpi_{\mathrm{min}}+\varpi_{\mathrm{max}})/2$
\STATE \textbf{If} $\varpi_j\mathbf{v}^{(m)}\in \mathcal{F}$, \textbf{then} set $\varpi_{\mathrm{min}}=\varpi_j$
\STATE \textbf{Else} set $\varpi_{\mathrm{max}}=\varpi_j$
\STATE $j=j+1$
\UNTIL $\varpi_{\mathrm{max}}-\varpi_{\mathrm{min}}\leq\delta_{\mathrm{BS}}$
\STATE Calculate projection $\bm{\pi}(\mathbf{v}^{(m)})=\varpi_{\mathrm{min}}\mathbf{v}^{(m)}$
\end{algorithmic}
\end{algorithm}
We note that the projection of vertex $\mathbf{v}^{(m)}$ on the upper boundary of set $\mathcal{F}$, i.e., $\bm{\pi}(\mathbf{v}^{(m)})$, is required in each iteration of \textbf{Algorithm 1}. For this purpose, we propose a bisection search-based algorithm to obtain the desired projection of $\mathbf{v}^{(m)}$, i.e., $\bm{\pi}(\mathbf{v}^{(m)})=\varpi\mathbf{v}^{(m)}$, where $0<\varpi<1$ is the projection parameter. The proposed algorithm is summarized in \textbf{Algorithm 2}. Next, we interpret the key step of \textbf{Algorithm 2}, i.e., step 4. In particular, in the $j$-th iteration, for a given projection parameter $\varpi_j$ and vertex $\mathbf{v}^{(m)}$, we have $\varpi\mathbf{v}^{(m)}\in\mathcal{F}$ if the following optimization problem is feasible
\begin{eqnarray}
\label{prob3}
&&\hspace*{-6mm} \mathrm{Find} \hspace*{6mm}\left \{\mathbf{W}_k,\mathbf{Z},\zeta_{\mathrm{II}},\tau_k,\kappa_k,\varrho_k\right \}\notag\\
&&\hspace*{-10mm}\subto\hspace*{1mm}\widetilde{\mbox{C1}},\widetilde{\mbox{C2}}, \widetilde{\widetilde{\mbox{C3a}}}, \widetilde{\widetilde{\mbox{C3b}}},\widetilde{\mbox{C5}},
\notag\\
&&\hspace*{8mm}\mbox{C6:}\hspace*{1mm} f_k(\mathbf{W},\mathbf{Z})-\varpi_jv_k^{(m)}g_k(\mathbf{W},\mathbf{Z})\geq 0,\hspace*{0.5mm}\forall k.
\end{eqnarray}
Now, the only barrier for efficiently solving problem \eqref{prob3} is the rank-one constraint $\widetilde{\mbox{C5}}$. To this end, we employ semidefinite relaxation to remove constraint $\widetilde{\mbox{C5}}$ from \eqref{prob3} and utilize CVX on the resulting convex problem. In addition, the tightness of the SDR is revealed in the following theorem.
\begin{theorem}
For a given positive constant $P^{\mathrm{max}}$ in the relaxed version of \eqref{prob3}, we can always obtain an optimal beamforming matrix $\mathbf{W}^*_k$ with unit-rank.
\end{theorem}
\textit{Proof:} The proof of \textbf{Theorem 1} follows similar steps as in \cite[Appendix A]{9183907} and thus is omitted here due to page limitation.
\par
\begin{algorithm}[t]
\caption{Two-Layer Optimization Algorithm}
\begin{algorithmic}[1]
\small
\STATE Set step size of time $\Delta T$ and iteration index $d=1$
\REPEAT
\STATE Generate a time allocation policy $\mathcal{T}_d=\left\{d\Delta T,\hspace*{1mm}T_{\mathrm{tot}}-d\Delta T\right\}$
\STATE Solve problem \eqref{prob3_1} for a given time $d\Delta T$
\STATE Based on the objective function value of problem \eqref{prob3_1}, obtain the maximum AoD uncertainty $\psi_{\mathrm{II}}$ in set $\Omega_{\theta|\xi}$, cf. \eqref{prob1_2}
\STATE For given $T_{\mathrm{tot}}-d\Delta T$ and $\psi_{\mathrm{II}}$, solve problem \eqref{prob2_3} by employing \textbf{Algorithm 1}
\STATE Store the solutions and the corresponding objective function values
\STATE $d=d+1$
\UNTIL $d\Delta T=T_{\mathrm{tot}}$
\STATE Select the optimal solution $\left\{\mathcal{T}_d, \mathbf{W}_k,\mathbf{Z},\zeta_{\mathrm{I}},\zeta_{\mathrm{II}}\right\}$ that corresponds to the maximum system sum rate
\end{algorithmic}
\end{algorithm}
\par
We propose a two-layer algorithm where the optimal time allocation policy is searched in the outer layer while the optimal beamforming strategies for both phases are obtained in the inner layer. The proposed two-layer algorithm is summarized in \textbf{Algorithm 3}. Some additional remarks on the proposed algorithms are as follows:
\par
\textit{i) Initial point:} We initialize \textbf{Algorithm 3} by setting the time allocation policy as $\mathcal{T}_1=\left\{\Delta T,\hspace*{1mm}T_{\mathrm{tot}}-\Delta T\right\}$, where $\Delta T\in\mathbb{R}$ is the step size and without loss of generality, we assume $T_{\mathrm{tot}}=D\Delta T$, where $D\in\mathbb{N}_+$. For \textbf{Algorithm 1}, to ensure the first polyblock $\mathcal{P}^{(1)}\supset\mathcal{F}$, we generate $\mathbf{v}^{(1)}$ by finding the utopia point for the considered problem \eqref{prob2_3}. In particular, the $k$-th term of $\mathbf{v}^{(1)}$, i.e., $v^{(1)}_k$, is given by $v^{(1)}_k=1+\frac{\mathrm{Tr}(\mathbf{G}_k)P^{\mathrm{max}}}{\sigma_{\mathrm{U}_k}^2}$.
\par
\textit{ii) Convergence and optimality:}
We note that in the $m$-th iteration of \textbf{Algorithm 1}, we have $\mathcal{P}^{(m)}\supset\mathcal{F}$. Recall that $\mathbf{v}^{(m)}$ and $\bm{\pi}(\mathbf{v}^{(m)})$ are the vertex of the polyblock $\mathcal{P}^{(m)}$ and the corresponding projection on the upper boundary of $\mathcal{F}$, respectively. As a result, the objective function value of $\mathbf{v}^{(m)}$, i.e., $\Phi\big(\mathbf{v}^{(m)}\big)$, is an upper bound of $\Phi\big(\bm{\pi}(\mathbf{v}^{(m)})\big)$. As \textbf{Algorithm 1} proceeds, we can progressively slash the difference between $\Phi\big(\mathbf{v}^{(m)}\big)$ and $\Phi\big(\bm{\pi}(\mathbf{v}^{(m)})\big)$. According to \cite{tuy2000monotonic}, for given $\epsilon_{\mathrm{PA}}\rightarrow 0$, \textbf{Algorithm 1} is guaranteed to converge to the globally optimal solution of \eqref{prob2_2}. A detailed proof about the convergence of \textbf{Algorithm 1} can be found in \cite[Theorem 1]{tuy2000monotonic}. Moreover, since problem \eqref{prob3_1} is a convex problem, we can efficiently obtain its optimum. In fact, the objective function value of \eqref{prob3_1} is monotonically decreases when increasing $T$. Also, for given time allocation policy $\mathcal{T}_d$ and resulting $\xi\big(\mathbf{R}_\mathrm{s}\hspace*{0.5mm}|\hspace*{0.5mm}\mathcal{T}_d\big)$ from \eqref{prob3_1}, we solve \eqref{prob3_1} to obtain the associated maximal system sum rate. Hence, as we implement the one-dimensional line search over $T$ in \textbf{Algorithm 3}, we actually complete the traversal over all possible time allocation policy and the corresponding CRLB value. Therefore, we can guarantee to obtain the optimal time allocation and beamforming policy that maximizes the system sum rate of the considered system for a maximum information leakage threshold.
\par
\textit{iii) Complexity:}
In the outer layer of \textbf{Algorithm 3}, we implement a low-complexity line search. In the inner layer, we solve a semidefinite programming (SDP) problem and a non-convex problem in step 4 and 6, respectively. In particular, the computational complexity of solving SDP problem \eqref{prob3_1} is given by $\mathcal{O}\Big(\mathrm{log}_2(\frac{1}{\delta_{\mathrm{IP}}})N_{\mathrm{T}}^{3.5}\Big)$, where $\mathcal{O}\left ( \cdot  \right )$ is the big-O notation and $\delta_{\mathrm{IP}}>0$ is the convergence tolerance factor when employing the interior-point method to solve the SDP problem \eqref{prob3_1} \cite{yu2020irs}. While the computational complexity of \textbf{Algorithm 1} is given by $\mathcal{O}\Big(\mathrm{log}_2(\frac{1}{\epsilon_{\mathrm{PA}}})K^{I_{\mathrm{PA}}}\big[I_{\mathrm{BS}}\big((K+1)N_{\mathrm{T}}^{3.5}+(K+1)^2N_{\mathrm{T}}^{2.5}\big)\big]\Big)$, where $I_{\mathrm{PA}}$ and $I_{\mathrm{BS}}$ are the number of iterations required for \textbf{Algorithm 1} and \textbf{Algorithm 2} to converge, respectively. 
\section{Simulation Results}
\begin{table}[t]\vspace*{0mm}\caption{System simulation parameters.\vspace*{0mm}}\label{ISAC_parameters}\footnotesize
\newcommand{\tabincell}[2]{\begin{tabular}{@{}#1@{}}#2\end{tabular}}
\centering
\begin{tabular}{|l|l|l|}
\hline
    \hspace*{-1mm}$\sigma_{\mathrm{U}_k}^2$, $\sigma_{\mathrm{E}}^2$, $\sigma_{\mathrm{s}}^2$ & Noise power & $-90$ dBm \\
\hline
    \hspace*{-1mm}$T_\mathrm{s}$& Duration of the time slot & $1$ ms \\
\hline
    \hspace*{-1mm}$P^{\mathrm{max}}$ & DFRC BS maximum transmit power & $30$ dBm \cite{wei2018multibeam}\\
\hline
    \hspace*{-1mm}$C_k^{\mathrm{tol}}$ & Maximum tolerable information leakage & $0.5$ bits/s/Hz \\
\hline
    \hspace*{-1mm}$\Delta T$ & Step size of time & $0.1$ ms \\
\hline
    \hspace*{-1mm}$\epsilon_{\mathrm{PA}}$, $\delta_{\mathrm{BS}}$ & Convergence tolerance factors & $10^{-3}$ \\
\hline
\end{tabular}
\end{table}
In this section, we evaluate the performance of the proposed framework via simulations. In particular, we consider a multiuser ISAC system serving a $120$-degree sector of a circular cell with a radius of $200$ m. The DFRC BS is located at the center of the cell and is equipped with $N_{\mathrm{T}}=8$ antennas while $K=4$ users are uniformly and randomly distributed in the sector. The path loss exponent for the channels between the DFRC BS and  users is set to $3$ while that of the potential eavesdropper-involved channel is set to $2$. The path loss at the reference distance of $1$ m is set to $40$ dB \cite{9669263}. We assume that the small-scale fading coefficients of the BS-user links follow an independent and identically distributed Rayleigh distribution. The angular domain is equally separated into $L=120$ directions to generate the ideal beam pattern $\{P(\vartheta_l)\}_{l=1}^{L}$ for both phases. The beam pattern difference tolerance factors $\iota_{\mathrm{I}}$ and $\iota_{\mathrm{II}}$ are set to $0.1$, respectively. The upper bounds of AoD uncertainty before and after the first phase are set to $\widehat{\psi}=\frac{\pi}{6}$ and $\psi=3\sqrt{\xi}$, respectively. The bound of the distance uncertainty of the potential eavesdropper is assumed to be $D=10$ m. For ease of presentation, we normalize the maximum estimation errors for the coefficient of the round-trip channel according to $\varsigma=\frac{\left|\Lambda\right|}{\left|\overline{\Gamma}\right|}$ and set $\varsigma=0.1$. For comparison purposes, we also consider two baseline schemes. For baseline scheme 1, the available time is equally allocated to the two phases with duration $T=0.5T_{\mathrm{tot}}$ and no AN is employed in the second phase. For baseline scheme 2, we assume the ISAC system does not perform the first phase to refine the information of the potential eavesdropper, i.e., $T=0$. Then, we solve the problem in \eqref{prob1_2} based on the coarse knowledge about the potential eavesdropper. Unless otherwise specified, we adopt the parameters summarized in Table \ref{ISAC_parameters}.
\par
\begin{figure}[t]
\centering\includegraphics[width=3.4in]{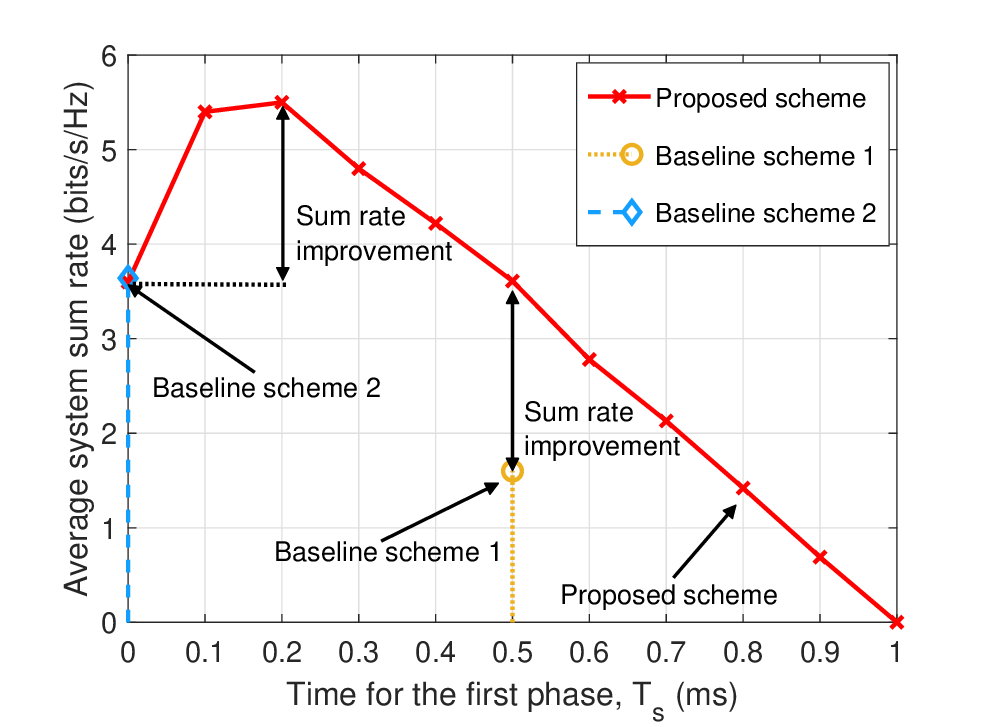}
\caption{Average system sum rate versus the time allocated to the first phase for different schemes.}\label{fig:time_rate}
\end{figure}
In Fig. \ref{fig:time_rate}, we investigate the average system sum rate versus the duration of the first phase, $T_\mathrm{s}$, for different schemes. In particular, we can observe that as $T_\mathrm{s}$ increases, the system sum rate of the proposed scheme first increases and then decreases. In fact, by increasing $T_\mathrm{s}$ from $0$ to $0.2$ ms, more time is allocated to refine the desired sensing information, thereby significantly reducing the CRLB for estimating the AoD of the potential eavesdropper. As a result, a much smaller AoD uncertainty is obtained at the beginning of the second phase. This allows the DFRC BS to execute a more precise beamforming policy, which in turn helps to improve the sum rate. However, as we further extend the duration of the first phase, only a shorter time is available for information transmission. In this case, the resulting average rate loss outweighs the gain induced by further reducing the AoD uncertainty that degrades the performance. Moreover, it can be seen from the figure that compared to the proposed scheme, both baseline schemes achieve dramatically lower system sum rate. Specifically, for baseline scheme 1, although a considerable fraction of the time slot is exploited to combat the AoD uncertainty of the potential eavesdropper, the system sum rate is still low due to the following two reasons. First, since AN is not employed, the DFRC BS has to reduce the transmit power for information beamforming for satisfying the maximum tolerable information leakage stated in constraint $\widetilde{\mbox{C3}}$. This inevitably results in lower systems sum rate. Second, as less time is exploited for information transmission, the average sum rate in the considered time slot also declines. As for baseline scheme 2, since the DFRC BS does not perform sensing to refine the AoD uncertainty, a large portion of transmit power has to be allocated to AN for security provisioning, leading to an inefficient information beamforming policy.
\par
\begin{figure}[t]
\centering\includegraphics[width=3.4in]{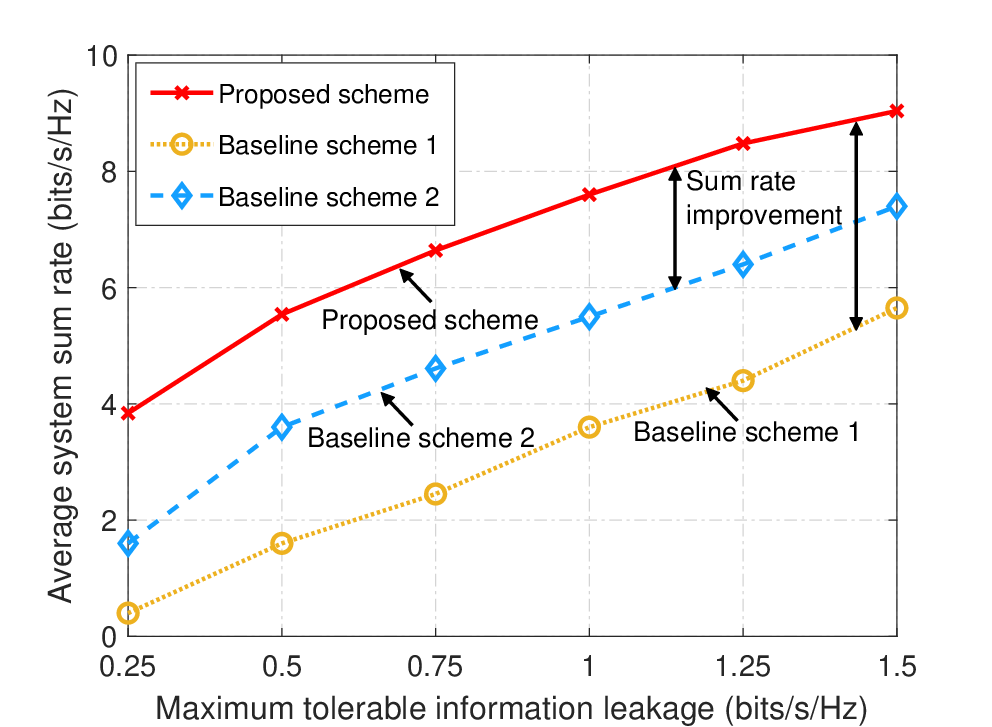}
\caption{Average system sum rate versus the maximum tolerable information leakage for different schemes.}\label{fig:leakage_rate}
\end{figure}
Fig. \ref{fig:leakage_rate} illustrates the average system sum rate versus the maximum tolerable information leakage $C_k^{\mathrm{tol}}$. As can be seen from the figure that for both the proposed scheme and two baseline schemes, the average system sum rate increases with $C_k^{\mathrm{tol}}$. However, the reasons behind this are different. For the proposed scheme, the gain originates from the following two reasons: First, a larger value of $C_k^{\mathrm{tol}}$ increases the tolerance of the DFRC BS against the AoD uncertainty. As a result, the DFRC BS is allowed to allocate less amount of time to the first phase to obtain a piece of less accurate AoD information about the potential eavesdropper, favoring a longer second phase for information transmission. Second, as $C_k^{\mathrm{tol}}$ increases, the DFRC BS is allowed to adopt a more aggressive power allocation policy where most of the available power is exploited for information transmission, as long as constraint $\widetilde{\mbox{C3}}$ is satisfied. In contrast, baseline scheme 1 can only benefit from the aforementioned second reason. As for baseline scheme 2, due to the fixed time allocation policy, the DFRC BS can only optimize the beamforming policy to boost the system performance, which loses the flexibility in striking the balance between the transmission time and AoD accuracy. This indicates the effectiveness of the proposed scheme by jointly optimizing all available resources in the considered system.
\section{Conclusion}
In this paper, we leveraged the sensing technique to enhance physical layer security for a multiuser communication system. In particular, we proposed a two-phase robust design framework, where the desired sensing information of the potential eavesdropper is acquired in the first phase while the secure transmission is performed in the second phase. Taking into account the uncertainties in the AoD, distance, and channel coefficient of the potential eavesdropper, we developed a monotonic optimization-based two-layer algorithm for determining the optimal time allocation and beamforming policy. Simulation results verified the remarkable ability of the proposed scheme to improve the physical layer security of wireless communication systems compared to the two baseline schemes. Moreover, our results revealed that a short-duration sensing phase can already efficiently lessen the CSI uncertainty of the potential eavesdropper, which helps combat wiretapping effectively.
\bibliographystyle{IEEEtran}
\bibliography{Reference_List}

\begin{thebibliography}{10}
\providecommand{\url}[1]{#1}
\csname url@samestyle\endcsname
\providecommand{\newblock}{\relax}
\providecommand{\bibinfo}[2]{#2}
\providecommand{\BIBentrySTDinterwordspacing}{\spaceskip=0pt\relax}
\providecommand{\BIBentryALTinterwordstretchfactor}{4}
\providecommand{\BIBentryALTinterwordspacing}{\spaceskip=\fontdimen2\font plus
\BIBentryALTinterwordstretchfactor\fontdimen3\font minus
  \fontdimen4\font\relax}
\providecommand{\BIBforeignlanguage}[2]{{%
\expandafter\ifx\csname l@#1\endcsname\relax
\typeout{** WARNING: IEEEtran.bst: No hyphenation pattern has been}%
\typeout{** loaded for the language `#1'. Using the pattern for}%
\typeout{** the default language instead.}%
\else
\language=\csname l@#1\endcsname
\fi
#2}}
\providecommand{\BIBdecl}{\relax}
\BIBdecl

\bibitem{6781609}
D.~W.~K. {Ng}, E.~S. {Lo}, and R.~{Schober}, ``Robust beamforming for secure
  communication in systems with wireless information and power transfer,''
  \emph{IEEE Trans. Wireless Commun.}, vol.~13, no.~8, pp. 4599--4615, Apr.
  2014.

\bibitem{wu2016secure}
Y.~Wu \emph{et~al.}, ``Secure massive {MIMO} transmission with an active
  eavesdropper,'' \emph{IEEE Trans. Inf. Theory}, vol.~62, no.~7, pp.
  3880--3900, Jul. 2016.

\bibitem{xu2019resource}
D.~Xu \emph{et~al.}, ``Resource allocation for secure {IRS}-assisted multiuser
  {MISO} systems,'' in \emph{Proc. IEEE Global Commun. Conf. (GLOBECOM)
  Wkshps}, Waikoloa, HI, USA, Dec. 2019, pp. 1--6.

\bibitem{gharavol2010robust}
E.~A. Gharavol \emph{et~al.}, ``Robust downlink beamforming in multiuser {MISO}
  cognitive radio networks with imperfect channel-state information,''
  \emph{IEEE Trans. Veh. Tech.}, vol.~59, no.~6, pp. 2852--2860, 2010.

\bibitem{skolnik2008radar}
M.~I. {Skolnik}, \emph{Radar Handbook}.\hskip 1em plus 0.5em minus 0.4em\relax
  McGraw-Hill Education, 2008.

\bibitem{9199556}
N.~{Su} \emph{et~al.}, ``Secure radar-communication systems with malicious
  targets: Integrating radar, communications and jamming functionalities,''
  \emph{IEEE Trans. Wireless Commun.}, vol.~20, no.~1, pp. 83--95, Jan. 2021.

\bibitem{9838753}
Z.~{Ren}, L.~{Qiu}, and J.~{Xu}, ``Optimal transmit beamforming for secrecy
  integrated sensing and communication,'' in \emph{Proc. IEEE Int. Conf.
  Commun. (ICC)}, Seoul, Korea, May 2022, pp. 5555--5560.

\bibitem{xu2022robust}
D.~{Xu} \emph{et~al.}, ``Robust and secure resource allocation for {ISAC}
  systems: A novel optimization framework for variable-length snapshots,''
  \emph{IEEE Trans. Commun.}, vol.~70, no.~12, pp. 8196--8214, Dec. 2022.

\bibitem{8999605}
F.~{Liu} \emph{et~al.}, ``Joint radar and communication design: Applications,
  state-of-the-art, and the road ahead,'' \emph{IEEE Trans. Commun.}, vol.~68,
  no.~6, pp. 3834--3862, Jun. 2020.

\bibitem{9746707}
Z.~Wei \emph{et~al.}, ``Safeguarding {UAV} networks through integrated sensing,
  jamming, and communications,'' in \emph{Proc. IEEE Int. Conf. Acoust.,
  Speech, Signal Process. (ICASSP)}, Singapore, May 2022, pp. 8737--8741.

\bibitem{9540344}
J.~A. {Zhang} \emph{et~al.}, ``An overview of signal processing techniques for
  joint communication and radar sensing,'' \emph{IEEE J. Sel. Topics Signal
  Process.}, vol.~15, no.~6, pp. 1295--1315, Nov. 2021.

\bibitem{8579200}
B.~{Tang} and J.~{Li}, ``Spectrally constrained {MIMO} radar waveform design
  based on mutual information,'' \emph{IEEE Trans. Signal Process.}, vol.~67,
  no.~3, pp. 821--834, Feb. 2019.

\bibitem{xu2022integrated}
D.~{Xu} \emph{et~al.}, ``Integrated sensing and communication in distributed
  antenna networks,'' \emph{accepted by IEEE Int. Conf. Commun. Wkshps. (ICC)},
  2023.

\bibitem{8288677}
F.~{Liu}, C.~{Masouros}, A.~{Li}, H.~{Sun}, and L.~{Hanzo}, ``{MU-MIMO}
  communications with {MIMO} radar: From co-existence to joint transmission,''
  \emph{IEEE Trans. Wireless Commun.}, vol.~17, no.~4, pp. 2755--2770, Apr.
  2018.

\bibitem{32276}
R.~Roy and T.~Kailath, ``{ESPRIT}-estimation of signal parameters via
  rotational invariance techniques,'' \emph{IEEE Trans. Acoust. Speech, and
  Signal Process.}, vol.~37, no.~7, pp. 984--995, Jul. 1989.

\bibitem{1143830}
R.~Schmidt, ``Multiple emitter location and signal parameter estimation,''
  \emph{IEEE Trans. Antennas Propag.}, vol.~34, no.~3, pp. 276--280, Mar. 1986.

\bibitem{8974403}
D.~{Xu} \emph{et~al.}, ``Multiuser {MISO} {UAV} communications in uncertain
  environments with no-fly zones: Robust trajectory and resource allocation
  design,'' \emph{IEEE Trans. Commun.}, vol.~68, no.~5, pp. 3153--3172, May
  2020.

\bibitem{1703855}
I.~Bekkerman and J.~Tabrikian, ``Target detection and localization using {MIMO}
  radars and sonars,'' \emph{IEEE Trans. Signal Process.}, vol.~54, no.~10, pp.
  3873--3883, Oct. 2006.

\bibitem{xu2023coordinatedisac}
D.~{Xu}, C.~{Liu}, S.~{Song}, and D.~W.~K. {Ng}, ``Integrated sensing and
  communication in coordinated cellular networks,'' \emph{accepted by IEEE
  Stat. Signal Process. Wkshps. (SSP)}, 2023.

\bibitem{6860253}
J.~{Xu}, L.~{Liu}, and R.~{Zhang}, ``Multiuser {MISO} beamforming for
  simultaneous wireless information and power transfer,'' \emph{IEEE Trans.
  Signal Process.}, vol.~62, no.~18, pp. 4798--4810, Sept. 2014.

\bibitem{boyd2004convex}
S.~Boyd and L.~Vandenberghe, \emph{Convex Optimization}.\hskip 1em plus 0.5em
  minus 0.4em\relax Cambridge University Press, 2004.

\bibitem{xunetworkisac}
Y.~{Xu} \emph{et~al.}, ``Joint {BS} selection, user association, and
  beamforming design for network integrated sensing and communication,''
  \emph{arXiv:2305.05265}, 2023.

\bibitem{ben2009robust}
A.~Ben-Tal, L.~El~Ghaoui, and A.~Nemirovski, \emph{Robust Optimization}.\hskip
  1em plus 0.5em minus 0.4em\relax Princeton University Press, 2009, vol.~28.

\bibitem{grant2008cvx}
M.~Grant and S.~Boyd, ``{CVX}: Matlab software for disciplined convex
  programming, version 2.1,'' \emph{http://cvxr.com/cvx}, Jan. 2020.

\bibitem{tuy2000monotonic}
H.~Tuy, ``Monotonic optimization: Problems and solution approaches,''
  \emph{SIAM Journal on Optimization}, vol.~11, no.~2, pp. 464--494, 2000.

\bibitem{9183907}
D.~{Xu} \emph{et~al.}, ``Resource allocation for {IRS}-assisted full-duplex
  cognitive radio systems,'' \emph{IEEE Trans. Commun.}, vol.~68, no.~12, pp.
  7376--7394, Dec. 2020.

\bibitem{yu2020irs}
X.~{Yu} \emph{et~al.}, ``{IRS}-assisted green communication systems: Provable
  convergence and robust optimization,'' \emph{IEEE Trans. Commun.}, vol.~69,
  no.~9, pp. 6313--6329, Sept. 2021.

\bibitem{wei2018multibeam}
Z.~Wei \emph{et~al.}, ``Multi-beam {NOMA} for hybrid mmwave systems,''
  \emph{IEEE Trans. Commun.}, vol.~67, no.~2, pp. 1705--1719, Feb. 2019.

\bibitem{9669263}
D.~{Xu} \emph{et~al.}, ``Optimal resource allocation design for large
  {IRS}-assisted {SWIPT} systems: A scalable optimization framework,''
  \emph{IEEE Trans. Commun.}, vol.~70, no.~2, pp. 1423--1441, Feb. 2022.

\end{thebibliography}
\end{document}